\documentclass[a4paper,11pt]{article}
\pdfoutput=1
\usepackage{graphicx}% Include figure files
\usepackage[utf8x]{inputenc}
\usepackage{dcolumn}% Align table columns on decimal point
\usepackage{amsmath}
\usepackage{amssymb}
\usepackage{dcolumn}% Align table columns on decimal point
\usepackage{bm}% bold math
\usepackage{enumerate}
\usepackage[affil-it]{authblk}
\usepackage[usenames,dvipsnames]{xcolor}
\usepackage{cite}

\makeatletter
\let\@fnsymbol\@arabic
\makeatother

\begin{document}

\title{Baryogenesis and Dark Matter with Vector-like Fermions}

\author{M. Fairbairn\thanks{malcolm.fairbairn@kcl.ac.uk}~}

\author{P. Grothaus\thanks{philipp.grothaus@kcl.ac.uk}}

\affil{Theoretical Particle Physics and Cosmology Group\\ Physics Department\\
King's College London\\ London WC2R 2LS\\ UK}

\date{\today}

\maketitle

\begin{abstract}
We show that vector-like fermions can act as the dark matter candidate in the universe
whilst also playing a crucial role in electroweak baryogenesis through contributing
to the barrier in the one-loop thermal scalar potential.  In order for the new fermions
to give rise to a strong first order phase transition, we show that one requires rather
large Yukawa couplings in the new sector, which are strongly constrained by electroweak
precision tests and perturbativity.  Strong couplings between the dark matter candidate
and the Higgs boson intuitively lead to small values of the relic density and problems with dark matter
direct detection bounds. Nevertheless, when considering the most general realisation of
the model, we find regions in the parameter space that respect all current
constraints and may explain both mysteries simultaneously.
\end{abstract}

\begin{section}{Introduction}

The recent discovery of the (Brout-Englert-Kibble-Guralnik-Hagen) Higgs boson at the 
LHC~\cite{Chatrchyan:2012ufa,Aad:2012tfa}
has led to a new surge in research concerning its possible implications for beyond the Standard Model physics.  

At the same time, we know from galaxy rotation curves, clusters, lensing and structure formation
both in galaxies and within the plasma before the last scattering surface that the Universe
seems to contain about five times as much dark matter (DM) as matter.  So far there is no information
from the LHC which elucidates the nature of this mysterious component.

Furthermore, even though there is evidence for CP-violation in the Standard Model 
(see e.g.~\cite{Christenson:1964fg,Aaij:2011in}), it is not clear what is the explanation for the fact that
there is more matter than anti-matter in the Universe.  While a source of CP-violation is
an essential ingredient for this asymmetry, we also require some form of out-of-equilibrium
physics in order to ensure that any baryon asymmetry is not washed out~\cite{Sakharov:1967dj}.  
Baryon number violating processes are
known to exist in the Standard Model~\cite{'tHooft:1976up,Kuzmin:1985mm},
but the question of how to achieve a strong enough deviation from thermal equilibrium is
still a matter of discussion. This can be achieved by heavy particles decaying out of 
equilibrium~\cite{Fukugita:1986hr,Luty:1992un}
or through strong first order electroweak phase transitions~\cite{Anderson:1991zb}.

At the time of writing the couplings of the new boson to the rest of the Standard Model
seem to confirm that it does indeed behave in the way that one would expect
from a Standard Model Higgs boson~\cite{Giardino:2013bma,Ellis:2013lra,Falkowski:2013dza}.
However, there is still some uncertainty with regards to the branching ratios for the decay channels of the Higgs, in particular,
the situation for the $h \rightarrow \gamma \gamma$ channel has not yet been clarified.
Whereas the ATLAS experiment still observes a
slight excess in that channel \cite{ATLAS:2013oma}, the CMS experiment finds a value
completely consistent with the Standard Model expectation \cite{Palmer:2013xza}.
It is clear that one of the key tasks in the immediate program of the LHC will
be to test these couplings as they may show evidence of coupling to other particles beyond the Standard Model.

One simple expansion of the Standard Model that could explain such a deviation is vector-like 
fermions, i.e. fermions whose left and right handed components transform in the same way under the Standard Model
gauge groups. Past works have demonstrated that not only can vector-like fermions enhance the diphoton
rate, but also that there are possible connections to other phenomena such as dark matter or baryogenesis. 
(For recent works on vector-like fermions, see for example~\cite{FileviezPerez:2011pt,Carena:2012xa,Joglekar:2012vc,ArkaniHamed:2012kq,
Almeida:2012bq,Batell:2012mj,Kearney:2012zi,Voloshin:2012tv,Arina:2012aj,Batell:2012zw,Davoudiasl:2012tu,Feng:2013mea,
Perez:2013nra,Joglekar:2013zya,Kyae:2013hda,Duerr:2013dza,Schwaller:2013hqa,Huo:2013fga,Garg:2013rba,Ishiwata:2013gma,Dermisek:2013gta,Choi:2013oaa}).
In this paper we will examine these connections in greater detail - in particular we aim to explore the compatibility of an explanation of both dark matter and baryogenesis
within one single set-up of vector-like fermions and see how future measurements in the
diphoton channel affect these links.

Given the value of the Higgs boson and the mass of the top quark, it appears
at the time of writing that the running of the quartic coupling of the Higgs
is such as to render the electroweak vacuum metastable~\cite{EliasMiro:2011aa,Degrassi:2012ry,Buttazzo:2013uya}.  
Generically at one loop,
bosons make the beta function of the quartic coupling more positive while
fermions make it more negative.  The addition of vector-like fermions to the
theory will consequently make this problem more acute and demands new physics to 
appear at some higher scale which we will not speculate on in this work. A possible solution would be 
supersymmetric versions of vector-like fermions, as investigated in for example~\cite{Joglekar:2013zya,Huo:2013fga,Feng:2013mea,Kyae:2013hda}.

The rest of the paper is organised as follows:
In section \ref{sec:lagrangian} we will present the Lagrangian and the mass matrices
of the model and write down the quantum numbers of the theory.  In section
\ref{method} we will describe the steps involved in looking for
parameters of the Lagrangian that provide viable dark matter candidates and a
strong electroweak phase transition without creating modifications to the propagation
of Standard Model particles which would have already been detected.  We then go on
to present the results in section \ref{results} before discussing our conclusions.

\end{section}

\begin{section}{Lagrangian and Mass Matrices \label{sec:lagrangian}}

In this work we investigate the model proposed in reference~\cite{Joglekar:2012vc}. We will use the
exact same notation here in order to avoid unnecessary confusion and present in this section the Lagrangian
and resulting mass matrices. Note that such a model may arise by gauging lepton number,
as explained in references~\cite{FileviezPerez:2011pt,Duerr:2013dza,Schwaller:2013hqa}.

The model consists of a SU(2) doublet $\ell_L'$
with Standard Model type couplings as well as corresponding singlets $e_R'$ and $\nu_R'$ plus, to make it
vector-like, mirrored fields with opposite chirality. See also table \ref{tab:qn} for clarification 
of the different fields and their quantum numbers. The Lagrangian is given by
\begin{eqnarray}
	{\cal L} &=& {\cal L}_{\rm SM}-m_\ell \bar{\ell}'_{\rm L}  {\ell }''_{\rm R} - m_e \bar{e}''_{\rm L} {e}'_{\rm R} 
	- m_\nu \bar{\nu}_{\rm L}'' {\nu}_{\rm R}' - \frac{1}{2}m' \overline{{\nu'_{\rm R}}^c} \nu'_{\rm R} 
	- \frac{1}{2}m'' \overline{{\nu''_{\rm L}}^c} \nu ''_{\rm L}\\
	&& -Y_c' (\bar{\ell}'_{\rm L} H ) {e}_{\rm R}' - Y_n' (\bar{\ell}'_{\rm L} \tau H^\dagger)  {\nu}_{\rm R}' 
	- Y_c'' (\bar{\ell}''_{\rm R} H ) {e}_{\rm L}'' - Y_n'' (\bar{\ell}''_{\rm R} \tau H^\dagger)  {\nu}_{\rm L}'' 
	+ \rm{h.c.} \nonumber
\label{lagrangian}
\end{eqnarray}
Here, the symbol $\tau$ represents the antisymmetric $2\times2$ matrix in SU(2) space with
non zero components $\tau_{12}=1$ and $\tau_{21}=-1$.
There are therefore nine free parameters of the model - five masses and four Yukawa couplings. 
\begin{table}[!t]
\begin{center}
\begin{tabular}{| l |c|c|c|}
\hline
Field & $\ell_{{\rm L}}',\ell_{{\rm R} }''$ & $e_{{\rm R}}', e_{\rm L}''$ & $\nu_{\rm{R}}', \nu_{\rm L}'' $ \\ \hline
${\rm SU(3)} \times {\rm SU(2)} \times {\rm U(1)}$ & (1,2,-1/2) & (1,1,-1) & (1,1,0) \\ \hline
\end{tabular}
\caption{The new fields and their ${\rm SU(3)} \times {\rm SU(2)} \times {\rm U(1)}$ quantum numbers. \label{tab:qn}}
\end{center}
\end{table}
With $v=246$~GeV one obtains the following mass matrices for the charged sector
\begin{align}
	\begin{pmatrix} \overline{{e}_{\rm L}' } ~  \overline{{e}_{\rm L}''} \end{pmatrix} {\cal M}_c \begin{pmatrix} {e}_{\rm R}' \\ {e}_{\rm R}'' \end{pmatrix} + {\rm h.c.}
	\qquad \text{with} \qquad {\cal M}_c = 
	\begin{pmatrix} \frac{Y_c' v}{\sqrt{2}} & m_\ell \\ m_e & \frac{Y_c'' v}{\sqrt{2}} \end{pmatrix},
\end{align}
and for the neutral sector
\begingroup
\fontsize{11pt}{11pt}
\begin{align}
	\frac{1}{2}\begin{pmatrix}
         \overline{\nu'_{\rm L}}~\overline{{\nu'_{\rm R}}^c}~\overline{{\nu''_{\rm R}}^c}~\overline{\nu''_{\rm L}}\end{pmatrix}{\cal M}_n\begin{pmatrix}
	{\nu'_{\rm L}}^c\\        
	\nu'_{\rm R}\\ 
        \nu''_{\rm R}\\
        {\nu''_{\rm L}}^c
	\end{pmatrix}+ {\rm h.c.}
	\qquad \rm{with} \qquad {\cal M}_n = 
	\begin{pmatrix}
	0 & \frac{Y'_n v}{\sqrt{2}}& m_\ell & 0 \\
	\frac{Y'_n v}{\sqrt{2}} & m' & 0 & m_\nu\\
        m_\ell & 0 & 0 & \frac{Y''_n v}{\sqrt{2}}\\
        0 & m_\nu & \frac{Y''_n v}{\sqrt{2}} & m''
	\end{pmatrix}.
\end{align}
\endgroup
These matrices are diagonalised such that we obtain the masses and mixings of the physical states. The
charged sector requires two independent unitary matrices $U_L$ and $U_R$, whereas one unitary matrix
$V$ is sufficient for the neutral sector.
In this analysis we will use the symbols $N_{1,2,3,4}$ and $E_{1,2}$ for the mass eigenstates of
the neutral and charged fermions, respectively. Their masses will be written as $M_{N_{1,2,3,4}}$ and
$M_{E_{1,2}}$ and their couplings to the Higgs boson as, e.g., $C^{N_1 N_1 h}$ or $C^{E_1 E_1 h}$.

Now we will go on to show how we investigate different numerical values for the
free parameters in the model.  Although there is a very large nine-dimensional parameter space,
this will be considerably reduced by requiring the model to not only give rise to a viable DM
candidate, but also to trigger a first order phase transition, to keep
modifications to the electroweak precision variables small (as observed)
and to give an excess in the diphoton rate.

\end{section}

\begin{section}{Methodology \label{method}}
In this section we will outline the procedure for the different tests we perform on a given set of 
parameters to see if they are compatible with dark matter, baryogenesis, electroweak precision 
variables etc.  In the next section we will present the results of this analysis.

\begin{subsection}{Parameter Ranges and Diagonalisation}

Since we investigate the complete accessible nine-dimensional parameter space,
we rely on the diagonalisation routines from the SLHAplus 
package~\cite{Belanger:2010st} to find the mass eigenstates 
and mixing matrices of the new charged and the neutral vector-like fermions.
For this we use Singular Value Decomposition and Takagi Decomposition respectively.
The resulting lightest neutral state will then be our dark matter
particle $N_1$ with a mass $m_{\rm DM}$.
As we expect the expressions for the masses and mixing elements to be too complicated to be illuminating, we perform the complete work without
considering analytic expressions.

Our parameter ranges are given as follows:
\begin{center}
\begin{tabular}{ccc}
 ~$m' \in [0, 4000]$ GeV,~ & ~$m'' \in [0, 4000]$ GeV, ~& ~ $m_\ell \in [0,4000]$ GeV, ~\\[1mm]
$m_\nu \in [0, 4000]$ GeV, & $m_e \in [0, 4000]$ GeV, &  $Y_c' \in [0,3.6]$\,,  \\[1mm]
~$Y_c'' \in [0,3.6]$ , ~& ~$Y_n'\in [0,3.6]$ , ~&  $Y_n'' \in [0,3.6]$\,.  ~\\[1mm]
\end{tabular}
\end{center}
The upper limits on the mass parameters are set to 4~TeV as we
do not expect any changes for even heavier mass parameters, as we would enter
the decoupling regime more and more. We implement perturbativity 
by setting an upper limit on the Yukawa couplings of 3.6. 

\end{subsection}

\begin{subsection}{Electroweak Constraints}
In order to ensure that the additional particles do not create problems for 
the successful predictions of the Standard Model, we need to check that the 
quantum corrections they induce upon the propagators of the W and Z bosons are not too large.  
These corrections are usually parameterised in the form of two parameters, $S$ and $T$,
if the new particles are heavier than half the $Z$-boson mass~\cite{Peskin:1991sw}.
The expressions for the $S$ and $T$ parameters for this model 
have been derived in~\cite{Joglekar:2012vc} and we refer to this work and~\cite{delAguila:2008pw,Cynolter:2008ea}
for more discussions on the oblique parameters with vector-like fermions.

The allowed ranges for a Higgs mass of 125~GeV are $S = 0.04 \pm 0.09$ and  $T=0.07 \pm 0.08$ 
and these errors are correlated with a coefficient of 0.88~\cite{Beringer:1900zz}, 
forming a diagonally oriented error ellipse in the $S$-$T$ plane.

As discussed in the literature, the contributions to $S$ and $T$ can be kept within the experimental limits
if the mass splitting between the components of the
doublets, $\Delta m = (y_c - y_n) v$, is small. Hence, if the charged Yukawa couplings
are large, the neutral Yukawa couplings must be large and vice versa. Also, the larger the 
vector-like masses are, the larger the mass splitting may be and the easier it is to respect the
experimental limits (although we have found exceptions to this general rule which we will talk about more in section~\ref{subsec:result_oblique}).

Absent signals of new charged states at the LHC create a mass limit on $E_1$.
Motivated by discussions in~\cite{Joglekar:2012vc,ArkaniHamed:2012kq,Joglekar:2013zya}
we apply a conservative lower limit on the lightest charged state of 110~GeV 
to ensure that current missing signals are consistent with our results, but remark that
a detailed investigation of collider constraints and signals lies beyond the scope of this paper.

In our model there is no mixing between the new vector-like fermions and the Standard Model
leptons. Such a set-up may arise from a gauged lepton number (as pointed out in~\cite{FileviezPerez:2011pt,Duerr:2013dza,Schwaller:2013hqa}) and 
automatically avoids constraints from flavor physics and stabilises our dark matter candidate.
For works that investigate models with mixing between a vector-like
fourth generation and the SM fermions, see e.g.~\cite{Kearney:2012zi,Ishiwata:2013gma,Dermisek:2013gta}.

\end{subsection}

\begin{subsection}{Dark Matter}
\label{subsec:dm}

The Lagrangian is expressed in the language of LanHEP~\cite{Semenov:2010qt} such that we can use
micrOMEGAs~\cite{Belanger:2010gh} to calculate the relic abundance, $\Omega \rm h^2$,
and the spin-independent WIMP-Nucleon cross section $\sigma^{\rm SI}$.

Results from LEP constrain the mass of an additional Majorana neutrino to be
above 39~GeV~\cite{Amsler:2008zzb}. Whereas this limit might in some cases be avoided due to the mixing 
of the DM candidate with singlet fields, it still serves as a good first estimate, because in these low mass regions
a DM-$Z$ coupling must be present for consistent relic abundance. Hence, we show this limit as a dotted, vertical
black line at 39~GeV in our plots.

Constraints arising from the dark matter sector include the
relic abundance, direct searches and the invisible branching ratio of the Higgs ${\rm BR}_{\rm inv}$.
We demand the relic density to 
lie in the range $[0.1134,0.1258]$ as has been measured by the 
Planck collaboration~\cite{Ade:2013zuv} and apply the
strong limits on $\sigma^{\rm SI}$ of the 
XENON100 collaboration~\cite{Aprile:2012nq}.
If the Higgs boson is kinematically allowed to decay into two dark matter
particles, a significant contribution to ${\rm BR}_{\rm inv}$
may arise, which is constrained by global fits as shown in references~\cite{Giardino:2013bma,Ellis:2013lra,Falkowski:2013dza}.
We will apply a rather strong limit in this work by demanding ${\rm BR}_{\rm inv} < 0.2$, but note
that there are theoretical uncertainties involved~\cite{Djouadi:2013qya}.

For further discussions of dark matter with vector-like fermions we refer the reader
to references~\cite{Joglekar:2012vc,Arina:2012aj,Perez:2013nra,Schwaller:2013hqa,Choi:2013oaa}.

\end{subsection}

\begin{subsection}{Baryogenesis}
\label{subsec:baryogenesis}
To find out whether the model can achieve a strong first order phase transition,
we evaluate the full free energy density $\mathcal{F}$ numerically. In its complete form it is given by
\begin{equation}
\label{freeenergy}
\mathcal{F}(\phi,T) = V_{\rm SM}\left(\phi\right) \pm
\sum_i g_i V(m^2_i(\phi))
+T^{4}\sum_i g_{i}I_{\mp }\left[m_{i}\left(\phi \right) /T\right]
/2\pi ^{2},
\end{equation} 
where the first term $V_{\rm SM}\left(\phi\right)= -\frac{1}{2}\mu^2 \phi^2 + \frac{1}{4} \lambda \phi^4$
is the tree-level Higgs potential, the second term is the one-loop contributions with a plus for bosons
and a minus sign for fermions, and the last term is the thermal corrections with the integrals $I_{\mp}$ defined below.
 $T$ is the temperature,
$g_i$ the degrees of freedom and $m(\phi)$ the 
field dependent masses. Both sums run over all Standard Model particles and the new
vector-like fermions. Only the $W$ and $Z$ bosons as well as the top quark are 
included from the SM particles, because other SM particles have negligible effects.

When using renormalisation conditions that do not change the tree level 
vacuum expectation value and the Higgs mass (which is fixed to 125~GeV in this work), the one-loop
contributions to the free-energy are given by
\begin{equation}
V(m^2(\phi))=\frac{1}{64\pi^2}\,m^4(\phi)\log m^2(\phi)+P(\phi),
\end{equation}
with the polynomial $P(\phi)$
\begin{equation}
P(\phi)=\frac{1}{2}\alpha\, \phi^2+\frac{1}{4}\beta\,\phi^4.
\end{equation}
Here, the coefficients are
\begin{eqnarray}
\alpha&=&\frac{1}{64\pi^2}\left\{\left(-3\,\frac{\omega \omega'}{v}+
\omega^{\prime\, 2}+\omega\omega''\right)\log\omega-\frac{3}{2}
\frac{\omega\omega'}{v}+\frac{3}{2}\omega^{\prime\, 2}+\frac{1}{2}
\omega\omega''\right\} ,\nonumber\\
\beta&=&\frac{1}{128\pi^2 v^2}\left\{2\left(\frac{\omega\omega'}{v}-
\omega^{\prime\, 2}-\omega\omega''\right)\log\omega
+\frac{\omega\omega'}{v}-3\omega^{\prime\, 2}-\omega\omega''\right\},
\end{eqnarray}
with
\begin{eqnarray}
\omega&=&m^2(v)~,\nonumber\\
\omega'&=&\left.\frac{dm^2(\phi)}{d\phi}\right|_{\phi=v},\nonumber \\
\omega''&=&\left.\frac{d^2m^2(\phi)}{d\phi^2}\right|_{\phi=v} .
\end{eqnarray}
The last part of equation~(\ref{freeenergy}), the thermal corrections, include the loop functions
\begin{equation}
I_{\mp }\left( x\right) =\pm \int_{0}^{\infty }dy\, y^{2}\log \left(
1\mp e^{- \sqrt{y^{2}+x^{2}}}\right) ,
\end{equation}
where $I_{-}$ is the contribution from bosons and $I_{+}$ from fermions.

A first indicator whether a large enough deviation from thermal equilibrium exists for baryogenesis
to be possible is the ratio of the vacuum expectation value at the critical temperature, $v_c$,
divided by the critical temperature $T_c$ itself. If $v_c/T_c$ is larger than one, sphaleron processes in the broken
phase are suppressed and no wash-out of a baryon asymmetry can occur. In the rest of our
analysis this will form our criterion whether baryogenesis can be successful.\footnote{For 
concerns about gauge invariance of $v_c$ and $T_c$, see~\cite{Patel:2011th,Wainwright:2012zn}.}

Note, that baryogenesis also requires a new source of CP-violation, which could arise in this model 
from complex Yukawa couplings. We leave an investigation of this extended parameter space
open for future work and refer to reference~\cite{Voloshin:2012tv} for discussions on a possible signature
of these CP-phases in the diphoton decay channel of the Higgs.

It has been shown that fermions may induce a
strong first order phase transition~\cite{Grojean:2004xa,Carena:2004ha,Davoudiasl:2012tu}
and we refer the reader to these references for detailed discussions
about the involved mechanisms, but give a short summary here for completeness.

In~\cite{Grojean:2004xa} an effective field theory approach was used to
show that $\phi^6$-terms in the scalar potential can lead to a barrier at zero temperature. Whereas
they used a scalar singlet as an example to generate these terms, one can also think of $\phi^6$-terms arising from
fermions, such that this mechanism is applicable to fermionic extensions of the Standard
Model as well. However, we do not observe a zero temperature barrier in our case, concluding
that temperature corrections are crucial.

The mechanism in~\cite{Carena:2004ha} relies on decoupling of heavy
fermions from the plasma when they enter the broken phase resulting in a delay of the phase
transition towards cooler temperatures and enhancing its strength in this way.
Since some of our particles become lighter when they enter the broken phase,
this effect is not the driving force behind our barrier.

The main effect occuring here is described in~\cite{Davoudiasl:2012tu}.
Temperature corrections can drive the effectiv quartic coupling $\lambda_{\rm eff}$ negative while 
$\mu_{\rm eff}$ becomes positive, resulting in a barrier at finite temperature. 
A positive $\phi^6$ term stabilises the potential up to some energy scale.

Large Yukawa couplings necessary for the barrier intuitively stand
in conflict with the oblique parameters and the dark matter sector, because large
Yukawa couplings may quickly lead to an underproduction of dark matter and to
a WIMP-Nucleon cross section above current exclusion limits. Hence, 
a detailed discussion of the model is necessary and we
investigate in the following the compatibility of those constraints all together.

We used existing works in the literature and Mathematica to
cross-check our numerical results.
\end{subsection}
\begin{subsection}{Diphoton Rate}
\label{subsec:diphoton}

Since vector-like fermions don't change the production channels of the Higgs,
the excess in the Higgs diphoton channel is characterised by the ratio of 
the decay rates only:
\begin{equation}
	.
\end{equation}
The known result for the decay rate
\begin{align}
	\Gamma(h \to \gamma\gamma) \propto \left| A_1(\tau_W) + \frac{4}{3}A_{1/2}(\tau_t) + \frac{C^{E_1 E_1 h} v}{\sqrt{2} M_{E_1}}A_{1/2}(\tau_{E_1}) + \frac{C^{E_2 E_2 h }v}{ \sqrt{2}M_{E_2}}A_{1/2}(\tau_{E_2}) \right|^2 ,
\end{align}
includes $\tau_x=\frac{m_h^2}{4m_x^2}$ as the argument in the loop functions
\begin{equation}
A_{1/2} (\tau)  = \frac{2\left(\tau+\left(\tau-1\right) f \left(\tau\right)\right)}{\tau^2}~,
\end{equation}
\begin{equation}
A_1(\tau) =-2-\frac{3}{\tau}-\frac{3\left(2\tau-1\right) f \left(\tau \right)} {\tau^2 }~,
\end{equation}
for fermions and bosons, respectively. The function $f(\tau)$ is given by
\begin{align}
f\left(\tau\right) &= \begin{cases} \arcsin^2\sqrt{\tau}  & \text{for } \tau\leq1\,,\\
\displaystyle
-\frac{1}{4}\left(-i\pi+\log\left[\frac{1+\sqrt{1-\tau^{-1} }}{1-\sqrt{1-\tau^{-1}}}\right]\right)^2\quad \quad &\text{for } \tau>1\,.
\end{cases}
\end{align}
To show the general dependence of the diphoton rate on the parameters, we
show for completeness in figure~\ref{fig:diphoton} how $R_{\gamma \gamma}$ behaves
under a variation of $m_{\ell}$ and $y_c'$. For a fixed $m_{\ell}$ the diphoton rate becomes stronger
with an increasing Yukawa coupling, and for a fixed Yukawa coupling
$R_{\gamma\gamma}$ decreases while $m_{\ell}$ increases.
These are the two general trends which have been discussed in the
literature, see e.g.~\cite{Carena:2012xa,Joglekar:2012vc,ArkaniHamed:2012kq,
Almeida:2012bq,Batell:2012mj,Kearney:2012zi,Voloshin:2012tv,Batell:2012zw,Feng:2013mea,
Dermisek:2013gta,Garg:2013rba} for more discussions.
In everything that follows all accepted scenarios have an enhanced Higgs diphoton branching ratio
between 1 and 2.

We will now go on to present the effect of fulfilling these various criteria upon the interesting parameter values.
\end{subsection}
\end{section}
\begin{section}{Results\label{results}}
Here we will present the application of the procedures outlined in the previous section and show how the different criteria affect the parameter space of out model.
\begin{subsection}{Numerical Approach}
To explore the parameter space we initially used a naive monte carlo scanning technique but we then moved
on to a basic version of a Metropolis Hastings Monte Carlo Markov chain (MCMC) to search for good 
parameters more efficiently, especially to find regions with a strong first order phase transition.
Note that we do not use the full power of the MCMC to obtain confidence levels
as there is not a very well defined notion of priors for the parameters within a Lagrangian.

There are some plots which are the result of naive parameter scans where we vary the parameters
randomly. Such plots (e.g. figure \ref{fig:mechanisms_sigma}) will have, for example, 
very few points which are not excluded by the XENON-100 bound on the WIMP-nucleon
cross section.  This told us that we had to scan the parameter space more intelligently such that we
implemented the MCMC technique. Including the XENON-100 bound as a fitness criteria,
the Markov chain hence concentrated on acceptable regions and the shape of the plots
changed significantly (see, e.g. figure \ref{fig:mechanisms_final}).

We do not claim to have made a completely comprehensive scan of the parameter space,
but we have gone to some effort within the ranges described above to look for parameter 
combinations which satisfy all of our requirements (dark matter, baryogenesis, not too large S and T parameters) 
which also have Yukawa couplings as small as possible, in order to retain perturbativity.

One relic of our slightly incomplete scanning procedure is that the careful reader may be able to spot some 
lines on the scatter plots which are not due to any physical mechanism, but rather due to the behaviour of one
of the MCMC chains as it explores the parameter range.  We have tried to identify which features are physical 
and which are random and hopefully made this distinction clear through drawing attention to real features using 
colour coding and careful wording in the figure captions.
\end{subsection}
\begin{subsection}{Electroweak Precision Oblique Parameters}
\label{subsec:result_oblique}
\begin{figure}[!t]
\begin{center}
\resizebox{0.75\columnwidth}{!}{\input{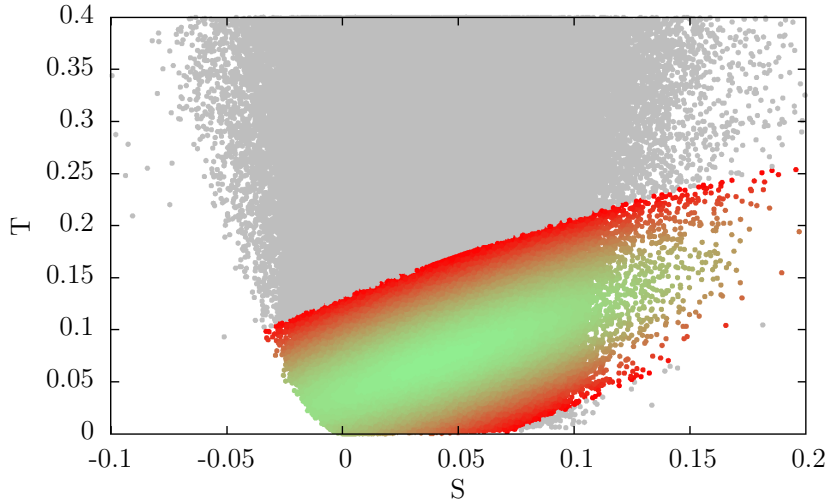}}
\caption{\it The 95\% confidence region of the oblique $S$ and $T$ parameters. Grey models which lie outside this region are rejected before any subsequent analysis.
\label{fig:oblique}}
\end{center}
\end{figure}
In figure~\ref{fig:oblique} we show the distribution of our 
models in the $S$-$T$ plane. The coloured region is part of the ellipse 
which represents the error on $S$ and $T$ and the correlation between those errors.
It can be seen that the models which arise from the Lagrangian~(\ref{lagrangian})
generally create larger modifications to the $T$ parameter rather than the $S$ parameter.
This is because the contribution of the mass splitting between the charged and neutral
components of the doublets, $\Delta m=(y_c -y_\nu )v$, comes in squared for $T$ but only 
enters the expression for $S$ within a logarithm~\cite{Peskin:1991sw,Joglekar:2012vc}. See also~\cite{delAguila:2008pw,Cynolter:2008ea}
for more discussions.

In some regions of the parameter space, the T parameter shows an unexpected behaviour. 
Whereas one would expect T to decrease when the vector-like mass terms are increased,
we find that in the most general case with non-zero neutral Yukawa couplings this does not necessarily have to 
be the case. To understand this better, we take a closer look at the equation for the 
T oblique parameter~\cite{Joglekar:2012vc}:
\begingroup
\fontsize{10}{12}
\begin{align}
4\pi s_{W}^2 c_{W}^2M_Z^2~T=&-2\sum_{j,k=1}^{2,4}\left(|U_{L_{1j}}|^2\left|V_{1k}\right|^2+|U_{R_{2j}}|^2\left|V_{3k}\right|^2\right)b_3\left(M_{N_k},M_{E_j},0\right)\notag\\
&+2\sum_{j,k=1}^{2,4}\,\rm{Re}\left(U_{L_{1j}}U_{R_{2j}}^*V_{1k}V_{3k}\right)M_{E_j}M_{N_k}b_0\left(M_{E_j},M_{N_k},0\right)\notag\\
&+\sum_{j,k=1}^4 \left(\left|V_{1j}\right|^2\left|V_{1k}\right|^2+\,\left|V_{3j}\right|^2\left|V_{3k}\right|^2\right)b_3\left(M_{N_j},M_{N_k},0\right)\notag\\
&-\sum_{j,k=1}^4\,\rm{Re}\left(V_{1j}V_{1k}^*V_{3j}V_{3k}^*\right)M_{N_j}M_{N_k}b_0\left(M_{N_j},M_{N_k},0\right)\notag\\
&+\left(|U_{L_{11}}|^4+|U_{R_{21}}|^4\right)M_{E_1}^2b_1\left(M_{E_1},M_{E_1},0\right)\notag\\
&+\left(|U_{L_{12}}|^4+|U_{R_{22}}|^4\right)M_{E_2}^2b_1\left(M_{E_2},M_{E_2},0\right)\notag\\
&+\left(2|U_{L_{11}}|^2|U_{L_{21}}|^2+2|U_{R_{12}}|^2|U_{R_{22}}|^2\right)b_3\left(M_{E_1},M_{E_2},0\right)\notag\\
&-\sum_{j,k=1}^2\rm{Re}\left(U_{L_{1j}}U_{L_{1k}}^*U_{R_{2j}}^*U_{R_{2k}}\right)M_{E_j}M_{E_k}b_0\left(M_{E_j},M_{E_k},0\right).
\end{align}
\endgroup
\begin{figure}[!t]
\begin{center}
\resizebox{0.65\columnwidth}{!}{\input{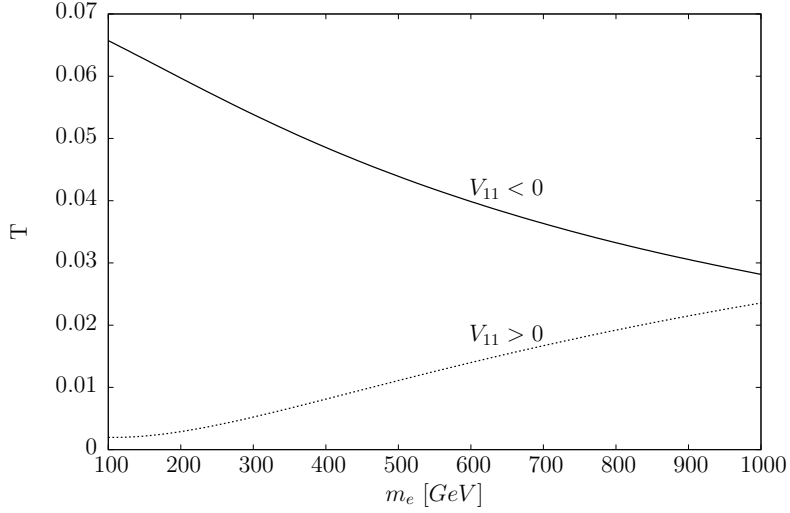}}
\caption{\it The T parameter and its dependence on the vector-like mass $m_e$ 
for different choices of the neutral Yukawa couplings leading to a change in sign of $V_{11}$.\label{fig:tpara}}
\end{center}
\end{figure}
Here, we have $s_W$ and $c_W$ as the sine and cosine of the Weinberg angle and $M_Z$ as the mass of the $Z$-boson.
The functions $b_0,b_1$ and $b_3$ are integrals defined as:
\begin{align}
b_0\left(M_1,M_2,q^2\right)&=\int_0^1\text{log}\left(\frac{\Delta}{\Lambda^2}\right)\,dx\,,\\
b_1\left(M_1,M_2,q^2\right)&=\int_0^1 x\,\text{log}\left(\frac{\Delta}{\Lambda^2}\right)\,dx\,,\\
b_3\left(M_1,M_2,0\right)&=\frac{M_2^2\,b_1\left(M_1,M_2,0\right)+M_1^2\,b_1\left(M_2,M_1,0\right)}{2}\,,\\
\Delta& = M_2^2x+M_1^2\left(1-x\right)-x\left(1-x\right)q^2.
\end{align}
We observe that the $2^{\rm nd}$ and $4^{\rm th}$ term depend explicitly on the sign of the
 elements of the neutral mixing matrix $V_{ij}$,
and consequently may increase or decrease the T-parameter. In figure~\ref{fig:tpara} we present 
the different behavior of T resulting
from a change in sign of the first component of the neutral mixing matrix $V_{11}$. For this 
figure we fixed the parameters as follows:
\begin{center}
\begin{tabular}{ccc}
$m' = 200$ GeV,~ & ~$m'' = 200$ GeV, ~& ~ $m_\ell = 750$ GeV, ~\\[1mm] 
$m_\nu = 100$ GeV, ~ & ~ $Y_c' = 1$,  ~&~ $Y_c'' = 1, $  ~ \\[1mm]
\end{tabular}
\end{center}
and chose $Y_n'=0.1,~Y_n''=0$ and $Y_n'=1,~Y_n''=1$ for the solid and dotted curve, respectively.
The first choice leads to a negative value for $V_{11}$ while the second one gives a positive $V_{11}$
while all other entries of the mixing matrix keep their signs unchanged. Overall, there is one additional
positive entry in $V_{ij}$ for the second possibility which leads to the shown dependence of the $T$ parameter
on $m_e$.

In all the plots in the paper following this subsection, scenarios lying outside the 95\% confidence
region have been removed from the data.
\end{subsection}

%%%%%%%%%%%%%%%%%%%%%%%%%%%%%%%%%%%%%%%%%%%%%%%%%%%%%%%%%%%%%%%%%%%%%%%%%%%%%%%%%%%%%%%%%%
\begin{subsection}{Dark Matter}
\label{subsec:result_darkmatter}
\begin{figure}[!t]
\begin{center}
\resizebox{0.75\columnwidth}{!}{\input{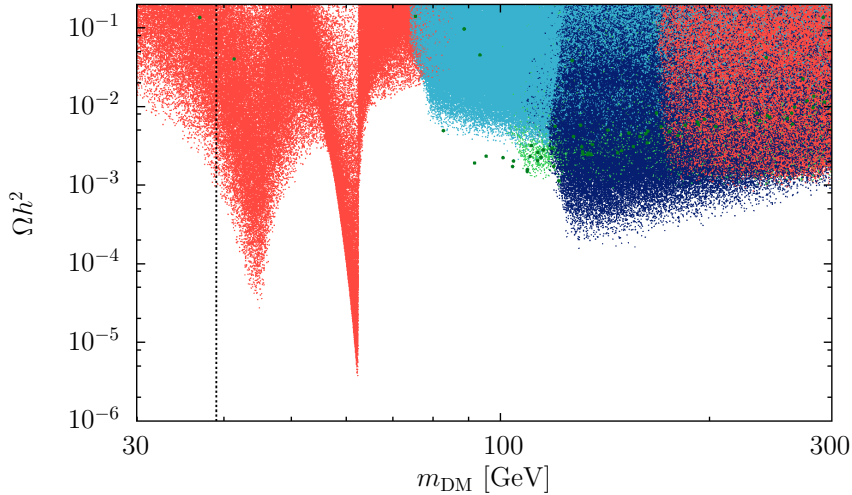}}
\caption{\it The relic density and the dominant annihilation channel responsible
 for relic abundance.\label{fig:mechanisms_omega}   
Red points represent annihilation into quark pairs, light blue points into 
gauge boson final states and dark blue points into a pair of Higgs 
bosons while light and dark green points denote co-annihilation with the 
lightest charged or next-to-lightest neutral fermion, respectively. In this and 
all following plots, the dotted vertical
line shows the LEP exclusion for an additional Majorana neutrino.}
\end{center}
\end{figure}
We extend existing discussions about dark matter in this model~\cite{Joglekar:2012vc}
by presenting results from an investigation of the complete parameter space performing
a simple monte carlo scanning.
 \begin{figure}[!t]
 \begin{center}
\resizebox{0.75\columnwidth}{!}{\input{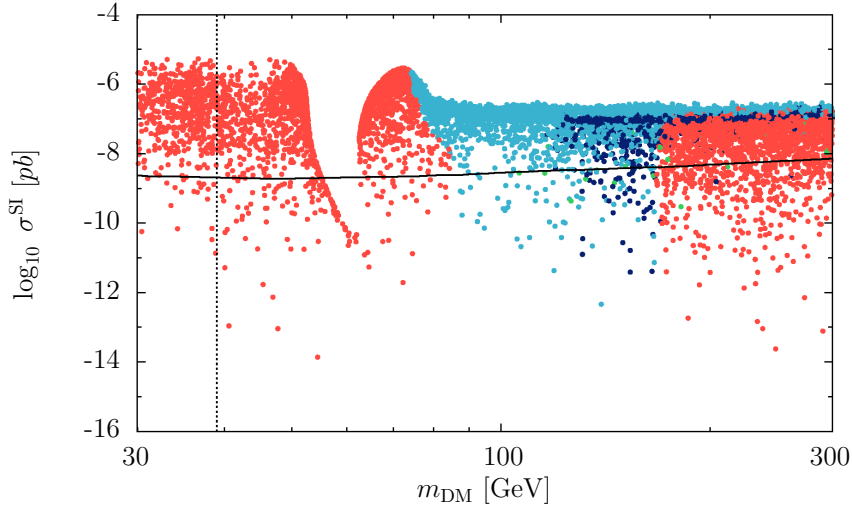}}
\caption{\it WIMP-Nucleon cross section $\sigma^{\rm SI}$ and the dark matter mass. 
All points give the correct dark matter relic abundance and are colour coded by the dominant
annihilation channel responsible for relic abundance.\label{fig:mechanisms_sigma}   
Red points represent annihilation into quark pairs, light blue points are gauge
boson final states, dark blue points denote Higgs pair production while light green points 
denote co-annihilation with the lightest charged state.
In this and all following plots the solid black line shows the XENON100 (2012) exclusion limit.}
 \end{center}
 \end{figure}
In figure~\ref{fig:mechanisms_omega} we show the relic abundance and the different annihilation
mechanisms that dominate during the freeze-out process. The shape is characterised by 
different resonances and thresholds. 

At a dark matter mass of about 45 GeV and at approximately 65 GeV
the $Z$- and $h$-resonances appear with a dominant decay into quark/antiquark pairs (mainly $b \bar{b}$),
represented by the red points. The red scenarios at masses above 170 GeV denote $t \bar{t}$ final states.
Above $m_{\rm DM} \approx 80$~GeV the possible annihilation into two $W$-bosons (and later also $Z$-bosons)
comes along with another decrease in the relic abundance (light blue points). A similar feature is observed at the Higgs threshold
at 125 GeV (dark blue points). Coannihilations with the lightest charged vector-like fermion $E_1$ (light green) set in at
approximately 100~GeV
while coannihilations with the second lightest neutral state $N_2$ (dark green) can be present at any mass range.
In this and all following plots the dotted, vertical black line indicates the mass limit 
from LEP for an additional Majorana neutrino, see section~\ref{subsec:dm} for more discussions on this.

We note here, that this picture can and will change when we concentrate on certain regions of the parameter space, as we
will do when we discuss baryogenesis. Coannihilations, which seem to under-produce dark matter here, will become
important as the dominant mechanism to set the relic abundance.

In figure~\ref{fig:mechanisms_sigma} we present the mapping of those scenarios that fulfill the relic
density condition (i.e. $\Omega h^2 \in [0.1134,0.1258]$) into the direct detection plane.  
We have again colour coded the diagram so it is easy to see the different sections of the diagram
where different mechanisms dominate the relic abundance calculation.
It is visible that direct detection limits can be avoided in the complete mass range. Since the scattering
process for the WIMP-Nucleon cross section is mediated by the Higgs, $\sigma^{\rm SI}$
indicates the coupling of our dark matter particle to the Higgs. This is best seen
at the $h$-resonance itself: To obtain the correct relic abundance, the resonantly enhanced annihilation rate asks for a 
a suppressed dark matter Higgs coupling, which translates into a suppressed $\sigma^{\rm SI}$.
\begin{figure}[!t]
\begin{center}
\resizebox{0.75\columnwidth}{!}{\input{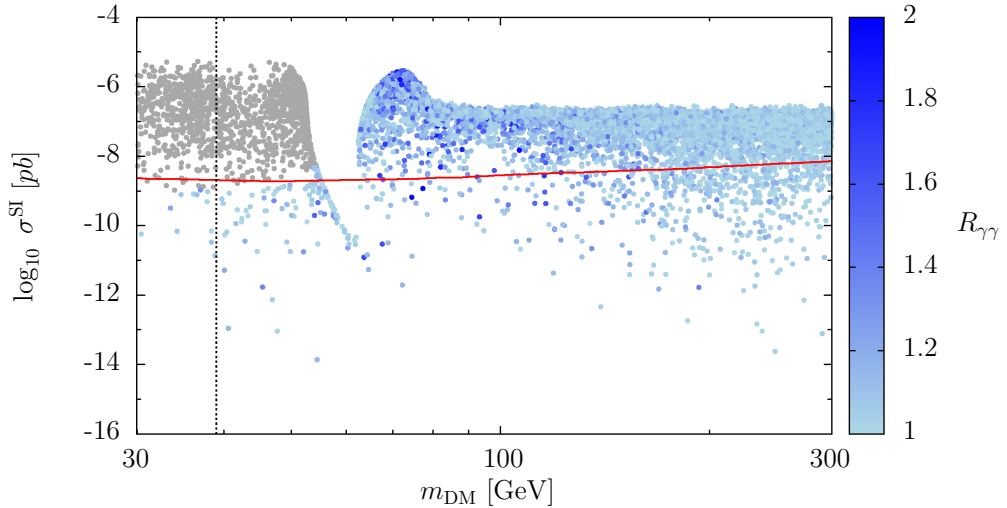}}
\caption{\it WIMP-Nucleon cross section $\sigma^{\rm SI}$ and the dark matter mass of a general scan.
The blue color indicates an enhanced Higgs to diphoton branching fraction. 
Grey points are excluded due to the invisible branching ratio of the Higgs.
\label{fig:sigma_diphoton}}
\end{center}
\end{figure}

The dark matter discussion is to some extent independent of an enhanced diphoton rate,
since the latter one is solely fixed by the charged sector. 
However, we note that the vector mass parameter for the doublet fields, $m_\ell$, relates
both sectors and can give rise to some correlations.

In figure~\ref{fig:sigma_diphoton} we show the diphoton rate in the direct detection plane
with a blue color coding and can observe a general trend of a decreasing excess in the diphoton channel
towards larger dark matter masses. This is because the charged states must become heavier
and their contribution to the diphoton rate gets suppressed as the dark matter mass increases.

In this plot we also apply the constraint from the invisible branching ratio of the Higgs boson.
Every scenario with ${\rm BR_{\rm inv}}>0.2$ is colored grey and therefore excluded. As expected,
these scenarios populate light dark matter masses with a large WIMP-Nucleon cross section.
If the LHC achieves a higher sensitivity for the invisible branching ratio, we can expect this
constraint to become more important for light dark matter masses and eventually be stronger
than direct searches.

Additionally, we cross-checked the predictions of the spin-dependent WIMP-Nucleon cross section $\sigma^{\rm SD}$
with the current limits from the XENON100 collaboration~\cite{Aprile:2013doa}, but found them to
be not constraining yet. Hence, we see that we find consistent realisations of the model at every accessible
dark matter mass.

\end{subsection}
%%%%%%%%%%%%%%%%%%%%%%%%%%%%%%%%%%%%%%%%%%%%%%%%%%%%%%%%%%%%%%%%%%%%%%%%%%%%%
\begin{subsection}{Baryogenesis}
\label{subsec:result_baryogenesis}
In order to work out if a given set of parameters leads to a strong first order
phase transition, we look at the free energy equation~(\ref{freeenergy}) and
iteratively vary the temperature until we get a phase transition.  
We then look for first order phase transitions and quantify the order 
parameter $v_c/T_c$ to see if sphaleron processes are suppressed at and below
the temperature of the phase transition, preventing 
washout of any baryon asymmetry.

We have analysed all the points we have found with good relic abundances
 in order to see which parameter sets give rise to a strong first order phase transition.
In figure~\ref{fig:direct_sfopt} we color code models
with a smooth cross-over or a weak phase transition light-grey, and those with $v_c/T_c>1$ 
dark-grey or in a red color scale. Dark-grey points are excluded by the invisible branching ratio of the Higgs boson,
whereas the red points fulfill all constraints considered in this work.

We can observe interesting behaviour at low dark matter masses.
There are two dips surrounding the $Z$-resonance and a thin band along the $h$-resonance.
This pattern is closely related to the constraints on the mixing of the dark matter
candidate arising from demanding that we can get a strong first order phase 
transition. In figure~\ref{fig:mixing} we show how strongly constrained the mixing
becomes once we apply the baryogenesis condition $v_c/T_c>1$. We show as an example
the second mixing element $V_{21}$ and note that similar plots exist for the other three
elements. Again, grey points satisfy electroweak and dark matter constraints and the red points
additionally the baryogenesis condition.
 \begin{figure}[!t]
\begin{center}
\resizebox{0.75\columnwidth}{!}{\input{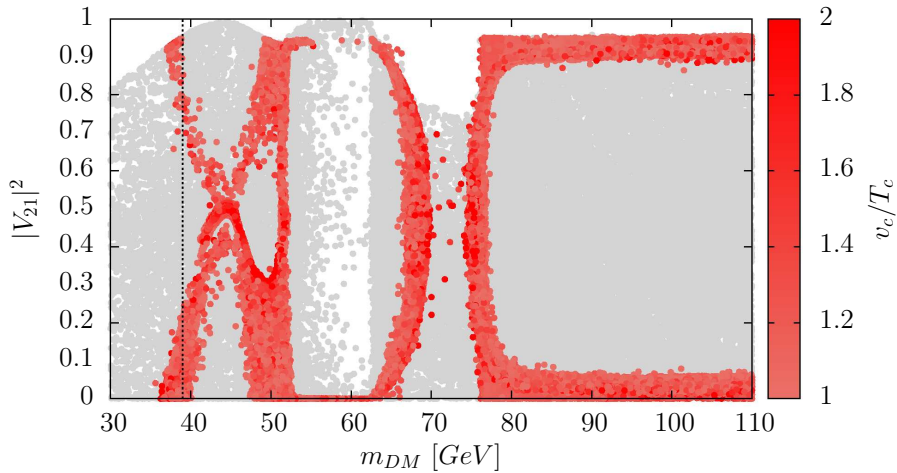}}
\caption{\it The second mixing element of the dark matter candidate
and the dark matter mass. Grey points fulfill electroweak and dark matter constraints, while
red points additionally give a strong first order phase transition.
\label{fig:mixing}}
\end{center}
\end{figure}

At the $Z$- resonance and at a mass around 70~GeV, one can observe that $|V_{21}|^2$ 
is always non-zero if we want to get a first order phase transition. As this holds for all the other mixing elements
 as well\footnote{Of course unitarity requests
some of them to be small, but around the $Z$-resonance they are still significantly further away from
zero than in other regions.}, we can understand why no suppression of $\sigma^{\rm SI}$ occurs.
Since the direct detection process proceeds via Higgs exchange, $\sigma^{\rm SI}$ will be proportional
to the Higgs coupling of the dark matter candidate, $C^{N_1hh}$, which is given by
\begin{align}
 C^{N_1hh}&= \frac{y_n'}{\sqrt{2}}~(V_{11}^* V_{21}^*+V_{11} V_{21})+ \frac{y_n''}{\sqrt{2}}~(V_{31}^* V_{41}^* + V_{31} V_{41}) \notag \\
&=\sqrt{2}~y_n' ~ {\rm Re}(V_{11} V_{21})+\sqrt{2}~ y_n'' ~ {\rm Re}(V_{31} V_{41})
\end{align}
For a suppressed direct detection rate, we generally need one of the mixing elements in each product to be suppressed
which cannot be achieved around the $Z$-resonance if we demand that the vector-like fermions are important for baryogenesis.

A second thing to note in figure~\ref{fig:direct_sfopt} is the tendency of smaller
order parameters towards larger dark matter masses. This is again explained through approaching
the decoupling limit. As all masses increase, the behavior of the free energy
becomes more and more standard model like and the first order phase transition
gets weaker and weaker. This puts an upper limit on the dark matter candidate of around 300~GeV.

In figure \ref{fig:mechanisms_final} we look at the same data as in figure \ref{fig:direct_sfopt} but this time
we colour code the annihilation channels. With this plot we can now understand the sharp
edge setting in slightly below 80~GeV in figure~\ref{fig:direct_sfopt}. There, coannihilations of the dark matter
particle with the second lightest neutral state into two $W$-bosons set in. The following dip at 125~GeV
can now be understood with the possible annihilations processes into two Higgs bosons. We note here again, that we do expect
paradoxes with figure~\ref{fig:mechanisms_omega} and~\ref{fig:mechanisms_sigma} where coannihilations seemed to
be unimportant for the relic abundance, because we changed our scanning procedure in such a way that we find 
interesting regions where baryogenesis is possible.
\begin{figure}[!t]
\begin{center}
\resizebox{0.75\columnwidth}{!}{\input{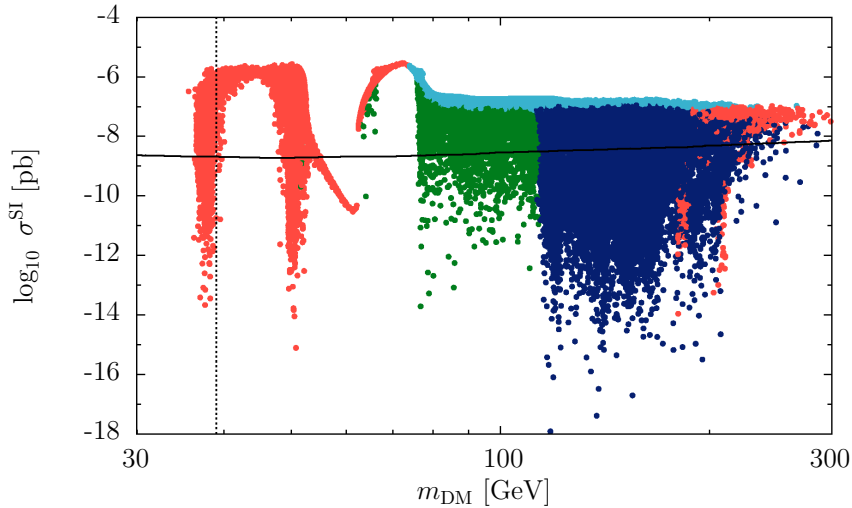}}
\caption{\it WIMP-Nucleon cross section $\sigma^{\rm SI}$ and the dark matter mass. 
All scenarios produce the correct dark matter relic abundance and give
a strong first order phase transition. They are colour coded
by the dominant annihilation channel responsible for relic abundance.\label{fig:mechanisms_final}   
Red points represent annihilation into quark pairs, light blue points are gauge
boson final states, dark blue points denote Higgs pair production and dark green points 
denote co-annihilation with the second lightest neutral state into a pair of quarks 
or $W$-bosons.}
\end{center}
\end{figure}

It is clear that we can obtain strong first order phase transitions
and good dark matter abundance whilst avoiding the XENON100
bound. There are regions at particular masses below 125 GeV and for a range of
masses above 125 GeV.  We would like to learn more about the characteristics
of the parameters which satisfy these criteria, but with nine free parameters
there are many combinations which can lead to good results. There is however
one feature of the parameter sets which gives rise to strong first order phase
transitions, and that is that they usually require large Yukawa couplings between
the Higgs and the additional fermions. In order to do that we look at a first benchmark
point (BM1), which has been chosen because it satisfies all the requirements we have
in terms of dark matter and baryogenesis with relatively small Yukawa
couplings (relative to other parameter sets that give us baryogenesis). See 
table~\ref{tbl:bm} for the numerical values of the parameters.
We then vary the Yukawa couplings to see the effect upon the strength
of the phase transition.

Figure~\ref{yukplot} shows the order of the phase transition
as a function of the primed charged and neutral Yukawa couplings.  While both Yukawa couplings increase the order
of the phase transition, it seems that for the benchmark point the neutral Yukawa
takes the lead role in realising baryogenesis. It becomes clear
that a large ratio $v_c/T_c$ only appears for large Yukawa couplings.

Related to this need for large Yukawa couplings, we point out that an enhanced diphoton rate
is unavoidable if the fermions should cause the phase transition to be strongly first order.
All simulated scenarios with $v_c/T_c>1$ have a minimal enhancement of $R_{\gamma \gamma}$
of 1.1. This implies that if ATLAS and CMS will with future measurements
agree on an enhanced diphoton rate, this could be interpreted as a direct
connection to baryogenesis. By the same token, if both experiments agree
on a Standard Model like decay rate, baryogenesis with vector-like fermions
in this set-up can be excluded.
 \begin{figure}[!t]
\begin{center}
\resizebox{0.75\columnwidth}{!}{\input{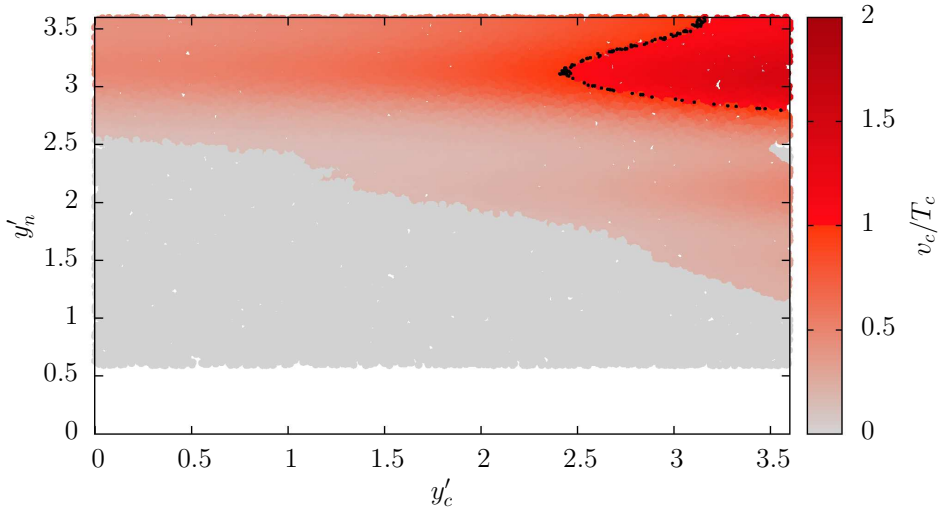}}
\caption{\it The order parameter and its dependence on the primed charged
and neutral Yukawa couplings. The black points have an order parameter of 1 and indicate
the switchover from a weak to a strong first order phase transition. All other parameter of our
benchmark point BM1 have been fixed. Note that as we vary the Yukawa couplings from the benchmark point,
the other constraints (S and T parameters, dark matter relic abundance, XENON100 limit
and diphoton bound) may be violated. \label{yukplot}}
\end{center}
\end{figure}
\newpage
\end{subsection}
\end{section}
\begin{subsection}{Vacuum Stability}
 \label{subsec:vacuum_stability}
\begin{table}[!h]
\begin{center}
    \begin{tabular}{| c | c | c | c | c | c | c | c | c | c | }
    \hline
     & $m'$~[GeV] & $m''$~[GeV] & $m_\ell$~[GeV] & $m_\nu$~[GeV] & $m_e$~[GeV] & $y_c'$ & $y_c''$& $y_n'$ & $y_n''$ \\ \hline
    BM1 & 97.70  & 90.90  & 2172.0 & 124.50 & 482.80 & 2.73 & 2.82  & 2.98 & 2.99\\ \hline
    BM2 & 869.53  & 236.57 & 2514.70 & 215.13 & 445.66 & 3.59 & 3.48 & 3.06 & 2.88 \\ \hline
    \end{tabular}
\caption{Parameters of two benchmark points considered in the model.\label{tbl:bm}}
\end{center}
\end{table}
The addition of exclusively new fermions to the SM renders the electroweak vacuum unstable at some scale $\Lambda$. If 
we define $V(v)=0$, the scale $\Lambda$ is set by the value of $\phi$ for which the potential turns negative,
opening up the possibility for a decay of the electroweak vacuum. We show in this section that this instability scale
lies above all the fermion mass eigenvalues by explicitly plotting the 
zero temperature scalar potential for two benchmark points, see figure~\ref{fig:pot}.

In section~\ref{subsec:baryogenesis} we chose a benchmark point with smallest possible 
Yukawa couplings consistent with all our constraints. As larger values of the Yukawa couplings 
will generally worsen the stability of the vacuum, we additionally present a consistent realisation with Yukawa couplings
very close to the perturbativity limit. In table~\ref{tbl:bm} we give their specific parameters and in table~\ref{tbl:masses}
the mass eigenvalues as well as the instability scale. For all accepted models the instability scale
is larger than the heaviest fermion mass. We note
that the stabilising state(s) must appear below the instability scale, but just
above the scale of the physics discussed in this paper.
\begin{figure}[!t]
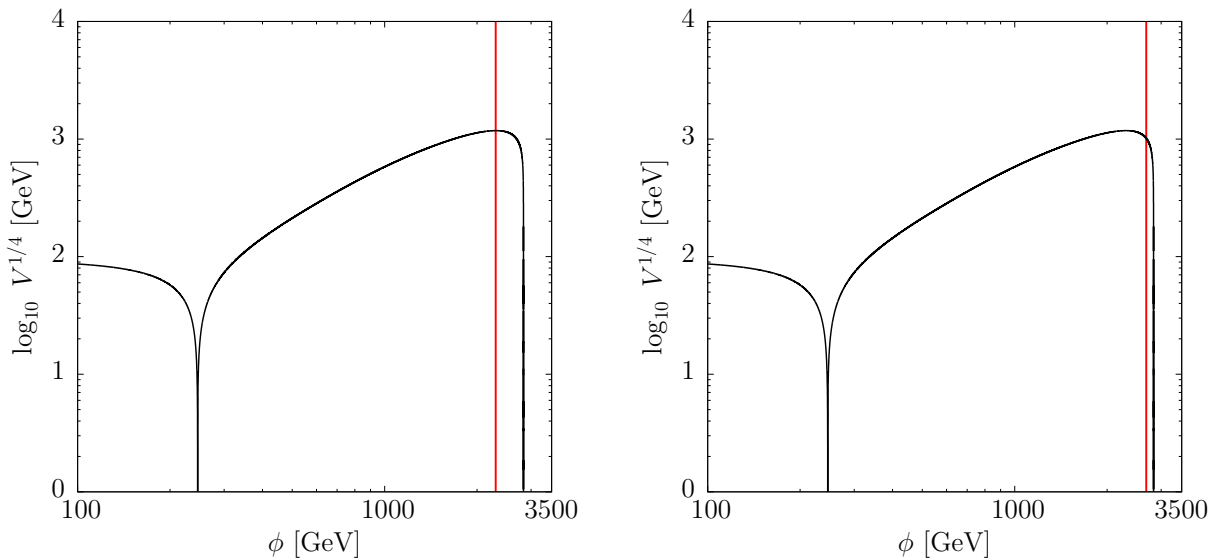

\begin{center}
 \begin{minipage}{0.49\textwidth}
\resizebox{.9\columnwidth}{!}{\input{pot_bm1}}
\end{minipage}
\hfill
\begin{minipage}{0.49\textwidth}
\resizebox{.9\columnwidth}{!}{\input{pot_bm2}}
\end{minipage}
\caption{\it The zero temperature scalar potential for benchmark point 1 
(left panel) and 2 (right panel). The scale of instability occurs above the largest
fermion mass eigenvalue, which is indicated by a red vertical line at the corresponding energy scale. \label{fig:pot}}
\end{center}
\end{figure}
 \begin{table}[!h]
 \begin{center}
     \begin{tabular}{| c | c | c | c | c | c | c | c | }
     \hline
      & $m_{N_1}$~[GeV] & $m_{N_2}$~[GeV] & $m_{N_3}$~[GeV] & $m_{N_4}$~[GeV] & $m_{E_1}$~[GeV] & $m_{E_2}$~[GeV] & $\Lambda$~[GeV] \\ \hline
     BM1 & 85.96  & 92.42  & 2291.24 & 2301.45 & 354.62 & 2300.23 & 2833.23 \\ \hline
     BM2 & 210.11 & 841.35 & 2605.85 & 2660.49 & 276.74 & 2683.68 & 2857.79\\ \hline
     \end{tabular}
 \caption{Masses and instability scale for both benchmark points. \label{tbl:masses}}
 \end{center}
 \end{table}
\end{subsection}
\begin{section}{Discussion and Conclusions}
In this paper we have investigated extensions of the standard model in the form of vector-like fermions.
As has been pointed out in work~\cite{Joglekar:2012vc}, there is the possibility to obtain a viable dark matter candidate
and the potential to increase the branching ratio of the Higgs boson decay into two photons, a channel
which will be measured in more detail when the LHC starts to run again.

In section~\ref{subsec:result_darkmatter} we presented our results of a general dark matter scan and saw that this
candidate can explain all the dark matter abundance in the universe in the complete 
mass region we investigated. We showed in figure~\ref{fig:mechanisms_omega} which mechanisms
are the most important ones for effective annihilation in the early universe and looked
at their mapping into the direct detection plane in figure~\ref{fig:mechanisms_sigma}. We 
saw that the invisible branching ratio of the Higgs can cut into the parameter space,
but is not yet sensitive enough to be more important than limits on the WIMP-Nucleon cross section $\sigma^{\rm SI}$.
Models exist where $\sigma^{\rm SI}$ is smaller than $10^{-50}$ cm$^2$ which presents a fresh challenge for the
next generation of dark matter direct detection experiments.

We have investigated the effect upon electroweak baryogenesis of these extensions to the Standard Model
in section~\ref{subsec:result_baryogenesis}
and showed that one can obtain a strong first order electroweak phase transition. Therefore, this kind of 
models enables the kind of out-of-equilibrium physics required to fulfill the Sakharov
conditions. We have noted that in order to obtain successful baryogenesis, one requires relatively large
Yukawa couplings between the new states and the Higgs, compare figure~\ref{yukplot}. These Yukawa
couplings can also serve as a possible new source of CP-violation.

If we demand that this extra sector provides both our dark matter and a strong first order 
electroweak phase transition simultaneously, there are quite large ranges of possible dark matter masses
between 39~GeV and 300~GeV with notable absences of models where the dark matter 
mass is around 45 GeV or 70 GeV, as can be seen in figure~\ref{fig:direct_sfopt} and~\ref{fig:diphoton_result}.
We hence find that both these problems of modern particle physics can be addressed within this single set-up.

Once the LHC is operational again following its current shutdown, new data 
will clarify the Higgs to diphoton rate which will place tight constraints upon 
this set of models.  It will be interesting to discover which regions of the parameter 
space will still be available if the constraints on this rate are increased.
\begin{figure}[!t]
\begin{center}
\resizebox{0.75\columnwidth}{!}{\input{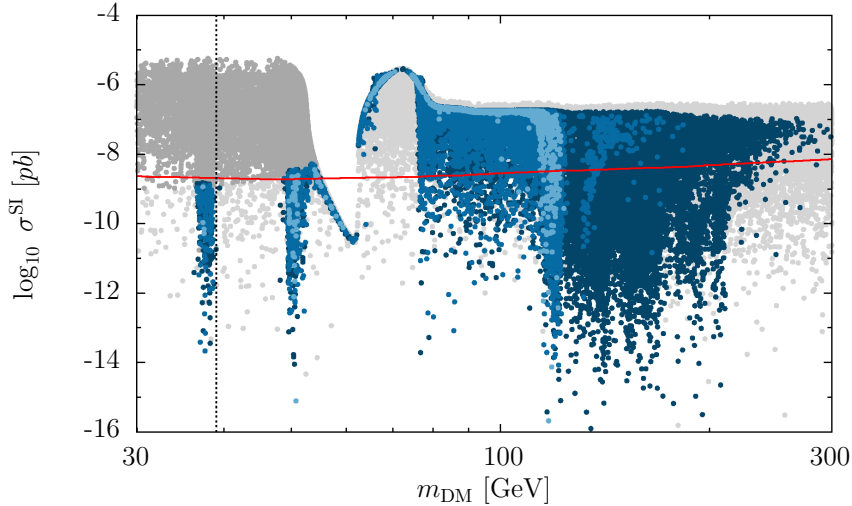}}
\caption{\it WIMP-Nucleon cross section $\sigma^{\rm SI}$ and the dark matter mass. 
The solid red line is the XENON100 (2012) limit, the dotted black line shows the mass
limit on a new Majorana neutrino by LEP.
All scenarios give the correct relic abundance and respect electroweak precision tests.
Dark grey points are, however, excluded by the invisible branching ratio of the Higgs.
The blue models respect all constraints, meaning they also posses a strong
first order phase transition.  They are coded in three blue tones
to indicate different upper limits on the diphoton excess $R_{\gamma \gamma}$ in order to show how future
measurements in the diphoton channel may affect our results. From dark to light, the 
corresponding upper limits on ${\rm R_{\gamma \gamma}}$ are 1.5, 1.3 and 1.2.
\label{fig:diphoton_result}   
}
\end{center}
\end{figure}

To explore this further, we show in figure~\ref{fig:diphoton_result} regions of models
that explain both dark matter and baryogenesis, colour coded according to the value of the excess in the diphoton channel.
As in the rest of the paper, we show a background of light-grey points that give 
a dark matter candidate fulfilling the relic density 
condition and electroweak constraints with an excess in the diphoton channel between 1 and 2.
Dark grey points are, however, ruled out by the upper limit on the invisible 
branching ratio of the Higgs. All blue points additionally give a strong
first order electroweak phase transition and form the scenarios consistent
with all our applied constraints.
The different shades of blue indicate different upper limits on
${\rm R}_{\gamma \gamma}$ ranking from 1.5 (darkest blue) via
1.3 (middle tone) to 1.2 (lightest blue). We can see that if in the future the 
excess in the diphoton channel consistently decreases in both experiments,
scenarios at high dark matter masses will be ruled out.
In those regions, all new vector-like fermions are relatively heavy and
for successful baryogenesis the charged Yukawa couplings need to be close to their largest
possible value still consistent with perturbativity, which comes
along with a relatively large enhancement in the diphoton decay channel.
Related to this we note again that all scenarios with a strong
first order phase transition show an enhancement of $R_{\gamma \gamma}$
of at least 1.1. This demonstrates how a better limit from the LHC in that channel can rule out electroweak baryogenesis within this model.

The 
best fitting values of the top quark mass and the Higgs mass suggest that the quartic 
coupling of the Higgs will become negative upon renormalisation to high energy scales, 
therefore rendering our current electroweak vacuum metastable \cite{EliasMiro:2011aa,Degrassi:2012ry,Buttazzo:2013uya}.  
The addition of fermions to the Standard Model in the way outlined in this paper will 
generically worsen this situation, such that the quartic coupling becomes negative at 
lower energies, potentially reducing the tunneling time from our vacuum to the true one to be comparable with the age of the Universe (as pointed out 
by the authors of~\cite{ArkaniHamed:2012kq}). We expect that the fact that
baryogenesis requires large couplings in this extended sector would make this 
potential problem worse. In section~\ref{subsec:vacuum_stability} we pointed out that
if the scenario set out in this work would turn out to be 
true, it would be a strong indication of additional particles at higher energies which 
would address this vacuum stability issue by adding positive contributions to the 
beta function for the quartic coupling.
\end{section}

\begin{section}{Acknowledgments}
We are happy to thank A.~Joglekar, P.~Schwaller, C.~Wagner and R.~Hogan for useful conversations.
MF is grateful for funding from the UK Science and Technology Facilities Council.
PG is supported by an ERC advanced Grant.
\end{section}

\end{document}